\documentclass[twocolumn,aps,prl,floatfix,superscriptaddress]{revtex4-1}
\usepackage{amsmath,amssymb}
\usepackage{graphicx,float}
\usepackage{times,txfonts}
\usepackage{color}
\usepackage{multirow}
\usepackage{bbold}
\usepackage{color}
\usepackage{soul}
\usepackage{hyperref}
\usepackage{times,txfonts}
\usepackage{natbib}
\usepackage{nicefrac}
\usepackage{blindtext}
\usepackage[export]{adjustbox}
\usepackage{mathrsfs}
\usepackage{textcomp}
\usepackage{blkarray}
\usepackage{multirow}

\newcommand{\abs}[1]{\left| #1 \right|}

\newcommand{\ket}[1]{\left| #1 \right\rangle}
\newcommand{\braket}[2]{\left\langle {#1{\left| \vphantom{#1 #2} \right.} #2} \right\rangle}

\renewcommand{\epsilon}{\varepsilon}

\def\VR{\kern-\arraycolsep\strut\vrule &\kern-\arraycolsep}
\def\vr{\kern-\arraycolsep & \kern-\arraycolsep}
\usepackage{sistyle}
\usepackage{soul}
\definecolor{lightblue}{RGB}{185,210,248}
\sethlcolor{lightblue}

\begin{document}

\title{Characterization of an underwater channel for quantum communications in the Ottawa River}

\author{Felix Hufnagel}
\affiliation{Department of Physics, University of Ottawa, 25 Templeton, Ottawa, Ontario, K1N 6N5, Canada}

\author{Alicia Sit}
\email{asit089@uottawa.ca}
\affiliation{Department of Physics, University of Ottawa, 25 Templeton, Ottawa, Ontario, K1N 6N5, Canada}

\author{Florence Grenapin}
\affiliation{Department of Physics, University of Ottawa, 25 Templeton, Ottawa, Ontario, K1N 6N5, Canada}

\author{Fr\'ed\'eric Bouchard}
\affiliation{Department of Physics, University of Ottawa, 25 Templeton, Ottawa, Ontario, K1N 6N5, Canada}

\author{Khabat Heshami}
\affiliation{Department of Physics, University of Ottawa, 25 Templeton, Ottawa, Ontario, K1N 6N5, Canada}
\affiliation{National Research Council of Canada, 100 Sussex Drive, Ottawa, Ontario K1A 0R6, Canada}

\author{Duncan England}
\affiliation{National Research Council of Canada, 100 Sussex Drive, Ottawa, Ontario K1A 0R6, Canada}

\author{Yingwen Zhang}
\affiliation{Department of Physics, University of Ottawa, 25 Templeton, Ottawa, Ontario, K1N 6N5, Canada}

\author{Benjamin J. Sussman}
\affiliation{Department of Physics, University of Ottawa, 25 Templeton, Ottawa, Ontario, K1N 6N5, Canada}
\affiliation{National Research Council of Canada, 100 Sussex Drive, Ottawa, Ontario K1A 0R6, Canada}

\author{Robert W. Boyd}
\affiliation{Department of Physics, University of Ottawa, 25 Templeton, Ottawa, Ontario, K1N 6N5, Canada}
\affiliation{Max Planck Institute for the Science of Light, Staudtstr. 2, D-91058 Erlangen, Germany}

\author{Gerd Leuchs}
\affiliation{Department of Physics, University of Ottawa, 25 Templeton, Ottawa, Ontario, K1N 6N5, Canada}
\affiliation{Max Planck Institute for the Science of Light, Staudtstr. 2, D-91058 Erlangen, Germany}

\author{Ebrahim Karimi}
\affiliation{Department of Physics, University of Ottawa, 25 Templeton, Ottawa, Ontario, K1N 6N5, Canada}
\affiliation{National Research Council of Canada, 100 Sussex Drive, Ottawa, Ontario K1A 0R6, Canada}
\affiliation{Max Planck Institute for the Science of Light, Staudtstr. 2, D-91058 Erlangen, Germany}

\begin{abstract}
We examine the propagation of optical beams possessing different polarization states and spatial modes through the Ottawa River in Canada. A Shack-Hartmann wavefront sensor is used to record the distorted beam's wavefront. The turbulence in the underwater channel is analysed, and associated Zernike coefficients are obtained in real-time. Finally, we explore the feasibility of transmitting polarization states as well as spatial modes through the underwater channel for applications in quantum cryptography.
\end{abstract}

\maketitle

\section{Introduction}


There are several different methods employed today for communicating underwater. The most widely used method is acoustic, capable of transmitting information over many kilometres~\cite{stojanovic:03}; however, the transmission rate is on the order of kilobits per second, limited by the speed of sound in water as well as the modulation rate of acoustic signals~\cite{heidemann:06}. A second method is to use radio-frequency (RF) signals, which can be easily incorporated into current communication networks. This technique is limited to communication distances on the order of several meters due to high absorption in water at radio frequencies. Both the acoustic and RF implementations suffer from the necessity of bulky and expensive equipment for both transmitting and receiving signals. Over the last decade, using the optical domain for underwater communication has gained increasing interest~\cite{wiener1980role,brock2009emerging}. With an optimal transmission window between blue and green (400-550\,nm) wavelengths, a propagation distance between 50-200\,m in clear water can be reached~\cite{zeng:17}. Higher data rates should additionally be achievable - up to gigabits per second depending on the scheme - allowing for larger data transfers and real-time communication. In~\cite{fasham:15}, a data rate of $20$~Mbps, at a distance of 200\,m, has been experimentally achieved.


In a realistic aquatic environment, there are several other factors beyond absorption which can limit the distance and quality of a marine optical communication link, i.e. scattering and turbulence. Scattering in water is dependent on the density and the size of particles in the channel and will contribute significantly to attenuation and therefore to the maximum achievable distance. Scattering is separated into two types: Mie scattering for particles on the order of the wavelength of the light, and Rayleigh scattering for particles much smaller than the wavelength~\cite{hulst:81}. Especially in water where there are relatively large plankton and mineral particles floating in the water, Mie scattering will have to be considered, along with Rayleigh scattering, due to the water molecules~\cite{mobley:94}. Another limiting factor when it comes to an actual implementation of an optical link is turbulence~\cite{nootz:16,nootz:17}. A spatially varying index of refraction from temperature and salinity differences through the optical link can result in beam wander as well as higher order distortion effects on the propagating beam. This can contribute both to loss and errors in the transmitted signal.
In optical communication, security---affected by factors such as errors in the channel---is an important feature for successful information transfer. Typically, a line-of-sight approach is implemented, making eavesdropping much more difficult, as opposed to the broadcasting method for acoustic and RF communication where the signal is sent in all directions. By considering quantum cryptographic schemes, the security can be further enhanced~\cite{Uhlmann:15}; for instance, quantum key distribution (QKD) allows authorized partners to communicate with unconditional security~\cite{bennett:84,gisin:02,scarani:09}. 

There are several different optical degrees of freedom which can be used to encode information in these QKD protocols. A popular option for direct line-of-sight channels is the polarization of photons, with successful experiments in free-space~\cite{schmitt:07,liao:17}. One limitation with polarization, however, is its inherently limited 2-dimensional Hilbert space, allowing for the maximum transmission of one bit per photon. 
The orbital angular momentum (OAM) degree of freedom of light, on the other hand, provides the potential of an unbounded state space, and thus unbounded encryption alphabet. Light beams carrying OAM possess a helical wavefront with $\ell$ intertwined helices, i.e. $\exp{(i\ell\phi)}$ where $\ell$ is an integer number and $\phi$ is the transverse azimuthal angle in polar coordinates~\cite{allen:92}. These beams possess a doughnut-shaped intensity profile due to the presence of a phase singularity at their centre ($\phi$ is undefined at the origin in cylindrical polar coordinates). The unbounded state space of these spatial modes allow us to implement high-dimensional quantum communication channels~\cite{mafu:13,mirhosseini:15,sit:17,bouchard:18b}, but they do come with unique challenges. One key challenge that has been observed with free-space communication is that turbulence in the channel can introduce errors in the transmitted information~\cite{krenn:14}. The measurement of OAM states is heavily dependent on the position of the incoming beam and thus these states are much more prone to errors from turbulence than polarization states which must just maintain their orientation. 

Underwater quantum communications have been numerically investigated~\cite{shi:15} and experimentally demonstrated in laboratory conditions using polarization~\cite{ji:17,zhao:19}, in outdoor conditions using the OAM degree of freedom~\cite{bouchard:18}, and over a 55\,m water channel using polarization~\cite{hu:18} and spatial modes~\cite{chen:19}. These experimental investigations have lead to several numerical investigations of QKD in underwater channels~\cite{tarantino:18,guo:18,gariano:19}.
In this Letter, we investigate the propagation of light through the Ottawa River in Canada's capital. In particular, we analyze the underwater turbulence by looking at the distorted wavefront and associated Zernike coefficients both obtained from a Shack-Hartmann wavefront sensor. Furthermore, we explore the transmission of polarization states of light and spatial modes of light through the underwater channel for quantum cryptography applications.

%



\section{Experiment}
The experiments presented here were conducted through the Ottawa River (latitude = 45.541048, longitude = -76.565719) during late August 2018. The water temperature was on average $20^{\circ}$~C for the duration of the experiment. However, the ambient temperature varied significantly from the middle of the day to the middle of the night. This contributed to turbulent conditions with water at the surface being heated or cooled more than the water below. Of course since it is a river, there were already naturally varying currents, which moved the water through the beams propagation path resulting in a changing index of refraction.

\begin{figure}[!t]
	\begin{center}
		\includegraphics[width=1.0\columnwidth]{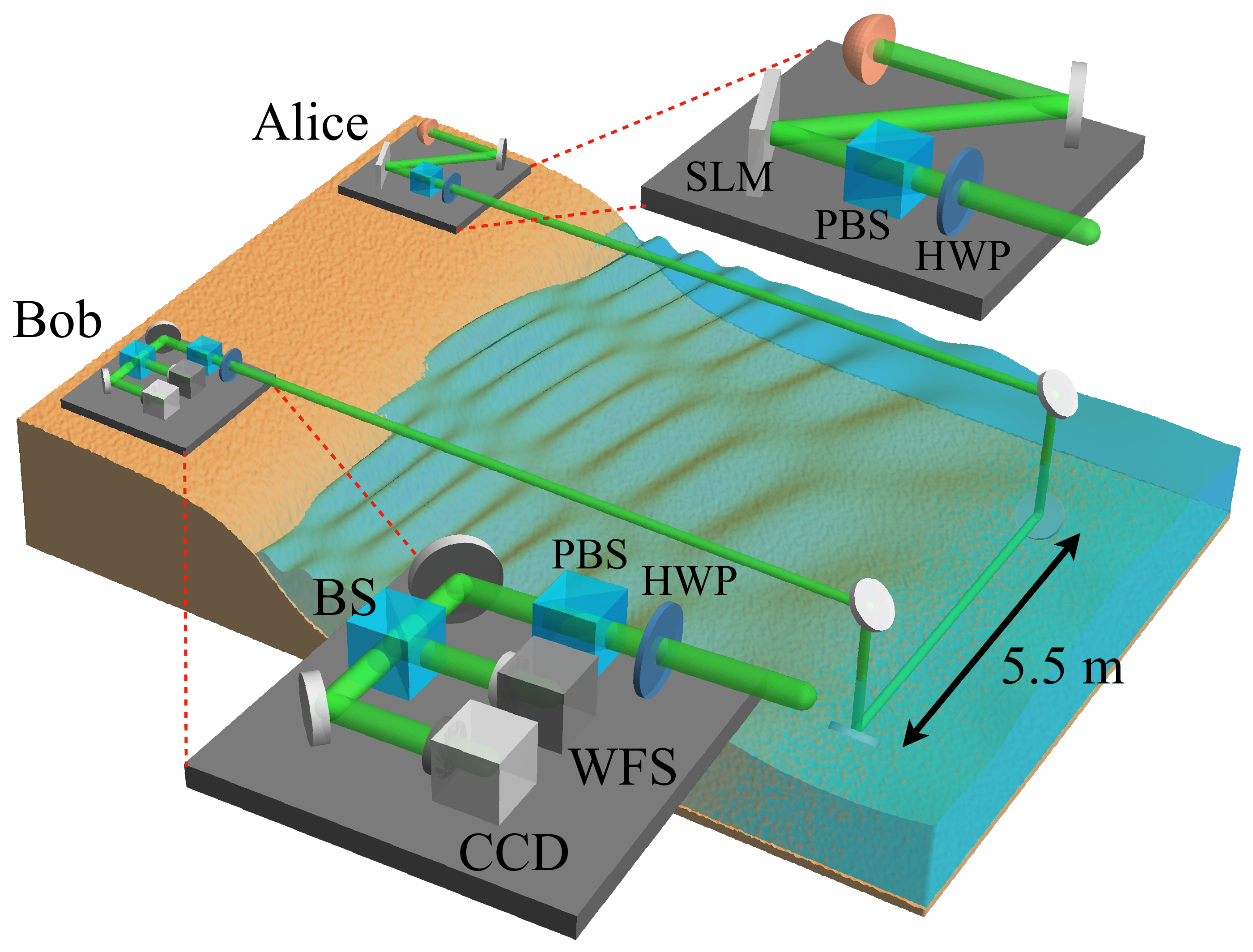}
		\caption[]{\textbf{Experimental setup.} Two breadboards positioned on the beach are used for the sender and receiver, Alice and Bob respectively. A CW laser at $\lambda=532$\,nm is sent to an SLM, polarizing beamsplitter (PBS), and half-wave plate (HWP) for state preparation at Alice's side of the link. This is then sent to a first periscope (composed of two mirrors) which brings the beam underwater, where it propagates to the second periscope 5.5\,m away. The receiver has a PBS and HWP for polarization measurements, and a beam splitter allows a CCD camera and Shack-Hartmann wavefront sensor (WFS) to take images.} 
		\label{fig:exp}
	\end{center}
\end{figure}

The results discussed in this work were taken using a 532\,nm laser diode. The sender and receiver units were mounted on breadboards along the shoreline of the river. As shown in Fig.~\ref{fig:exp}, the sender consisted of the laser, Spatial Light Modulator (SLM), and a half-wave plate. The wave-plates were used to prepare four linear polarization states, i.e. horizontal (H), vertical (V), anti-diagonal (A), and diagonal (D). The SLM is used for preparing the OAM states. This is done by displaying a phase hologram on the SLM and selecting the first diffracted mode from the hologram. The laser beam is then sent from Alice's breadboard on shore to the first periscope system that brings the beam underwater. The beam then propagates underwater, parallel to the beach, to the second periscope system where it is brought out of the water and sent to the receiver unit, see Fig.~\ref{fig:exp}. In order to eliminate air-water perturbations resulting from surface waves as the beam enters the water, a glass tube, closed at one end, is inserted within the periscope system to create an air-glass-water interface. At the receiver side, we captured the intensity of the beam for the polarization states of H, V, A, and D using a CCD camera. The camera only allows us to gain intensity information about the beam but not the phase. However in order to measure the phase of the beam, we place a Shack-Hartmann wavefront sensor (WFS) at the reciever. This device is made up of a micro-lens array placed in front of a CCD camera. The resulting effect is that the phase at each lens can be determined by the focus point of that lens on the CCD array. This allows one to determine the incidence angle of the given region of the beam and thus the phase relative to the rest of the beam. The accuracy of the wavefront sensor is limited by the number of micro-lenses in the array; the WFS that we use is the Thorlabs WFS20-7AR and has a $23\times23$ lenslet array with lenslet pitch of $150\,\mathrm{\mu}$m and focal length of 5.2\,mm.


Losses in the link due to scattering played a much larger roll in establishing a quantum channel than was expected. There has been analysis performed looking at the feasibility of quantum communication taking into account many factors including scattering~\cite{shi:15}. These studies, however, consider at worst the Jerlov Type III ocean water with scattering loss of 1.3\,dB$/$m. In our river channel, the total absorption was measured to be 5.4\,dB$/$m, significantly higher than even the worst water type considered in the previous calculations. This makes the absorption loss of 0.13\,dB$/$m for pure water negligible for practical considerations of achievable distance~\cite{smith:81}. Due to the large amount of scattering in the river, our experimental tests were limited to $\sim$5\,m. This high level of scattering was primarily due to large particles in the water ($d\gg\lambda$). The Mie scattering model is used when the particles diameter is on the same order as the wavelength of the light. This is typically for particles such as pollen, dust, and water droplets which are approximately the same size as the wavelength of the light. In our channel, since we were near the shore of the river, there was even larger visible plant matter and dirt floating in the water. This resulted in a large amount of light being absorbed or back reflected as opposed to being primarily forward scattered as in the regular Mie scattering regime. 


\begin{figure}[!htbp]
	\begin{center}
		\includegraphics[width=.8\columnwidth]{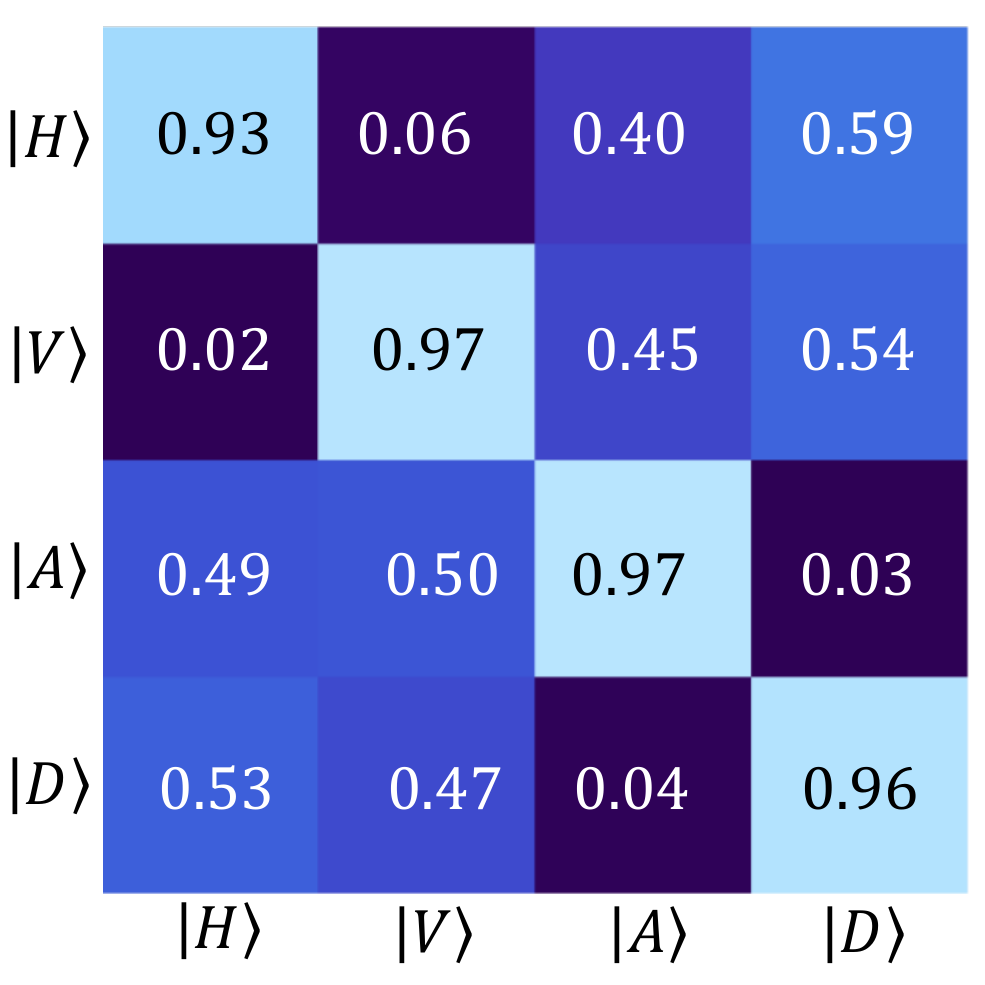}
		\caption[]{\textbf{Polarization probability-of-detection matrix.} The sender generates the linear polarization states of $\{\ket{H},\ket{V}\}$ or $\{\ket{A},\ket{D}\}$, chosen at random. The receiver randomly picks up one of the bases $\{\ket{H},\ket{V}\}$ or $\{\ket{A},\ket{D}\}$, and records the projection probability, whose numerical values are shown.}
		\label{fig:pol}
	\end{center}
\end{figure}

\begin{figure}[!htbp]
	\begin{center}
		\includegraphics[width=0.9\columnwidth]{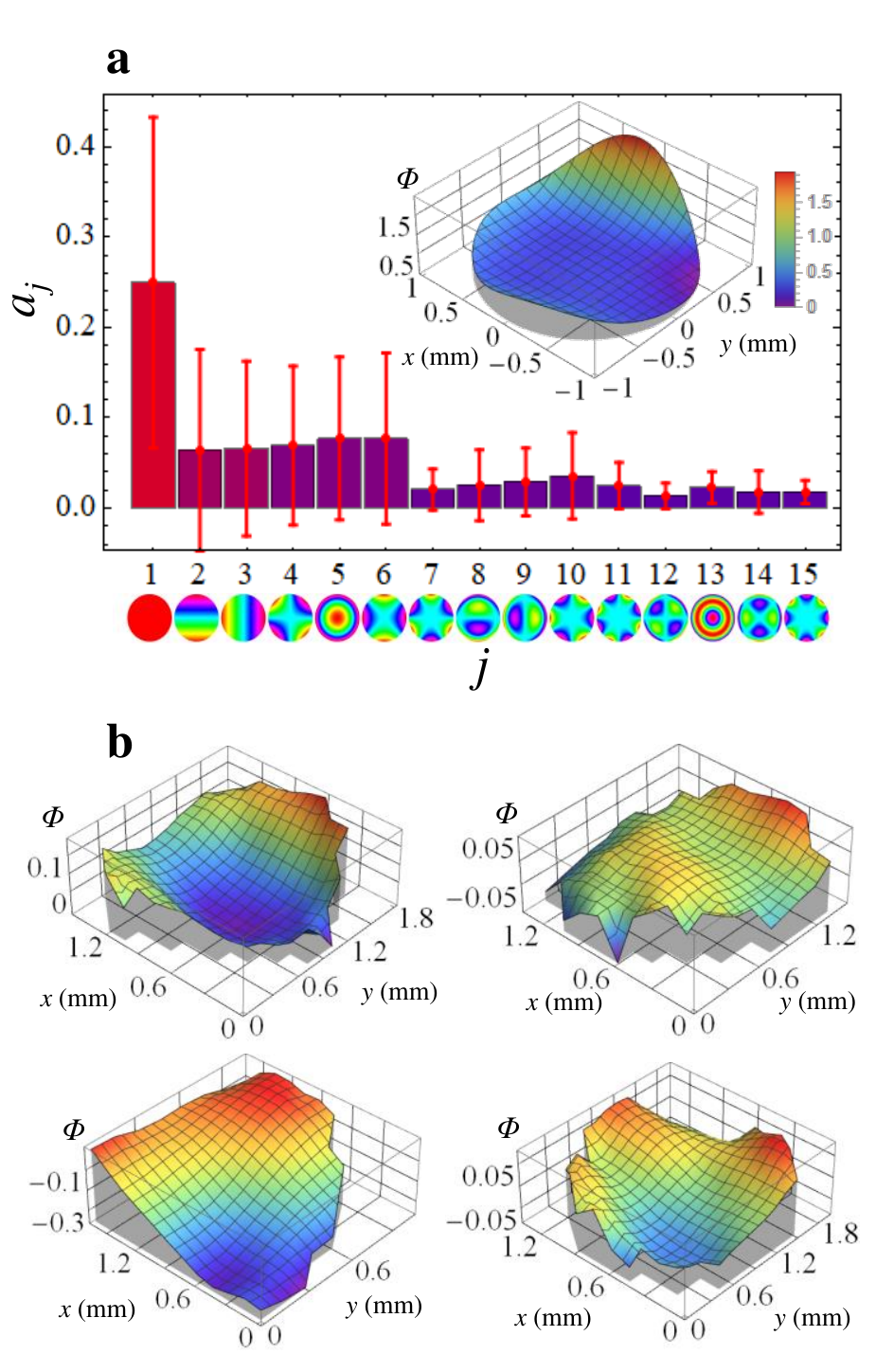}
		\caption[]{\textbf{Wavefront measurements}. A Gaussian beam is sent through the underwater channel to measure the wavefront. Plot \textbf{a} shows the average value for the magnitude of the Zernike coefficients taken from the wavefront sensor. The coefficients are separated by their radial degree corresponding to the color bars. of The inset is a plot of the wavefront given by these values. The plots in \textbf{b} are wavefronts measured at different times of a guassian beam through the 5\,m underwater link. The wavefront measurements are taken using an array of $150\,\mathrm{\mu}$m diameter lenses.}
		\label{fig:wf}
	\end{center}
\end{figure}

\section{Results and Discussion}
The first goal of this project was to establish that polarization QKD could be achieved in this highly turbulent and highly scattering channel. For the original BB84 protocol~\cite{bennett:84}, polarization states are chosen from a set of mutually unbiased bases (MUBs). We chose the bases to be $\ket{\psi_i}=\left\{ \ket{H}, \ket{V}\right\} $ and $\ket{\phi_i}=\left\{  \ket{A}, \ket{D}\right\} $. 
The defining property of MUBs is that a measurement in the correct basis reveals with certainty the state that the photon was in, while measurement in the wrong basis gives no information about the state of the photon, i.e. $\abs{\braket{\psi_i}{\phi_k}}^2 = 1/2$. Herein lies the security of QKD: an eavesdropper making a measurement in the wrong basis will be successful only 50\% of the time and will introduce errors when they are unsuccessful. The experimental probability-of-detection matrix for the polarization states is shown in Fig.~\ref{fig:pol}. The resultant error rate is 4.01~\%, which is below the threshold of 11.0~\% necessary to perform QKD with a 2-dimensional BB84 protocol. Though these results are obtained using classical light, single photons will behave in the same way so we can infer that an attempt done with a single photon system would be successful, and will result in a rate of $0.52$ bits per sifted photon. 


Although we achieved an error rate below the threshold, there are some residual errors in the system. In free space experiments, the errors are often attributed to optical turbulence, which comes from differences in the index of refraction along the path of propagation as described by the Kolmogorov theory of turbulence~\cite{kolmogorov:41}. Tip-tilt effects can result in beam wandering, while higher order effects can be present in high turbulence situations, resulting in distortion of the beam's profile~\cite{ren:16,bouchard:18}. The aberrations in the beam are often visible in the intensity of the beam; however, more precise information lies in the phase of the received beam.
\begin{figure*}[!htbp]
	\begin{center}
		\includegraphics[width=2.0\columnwidth]{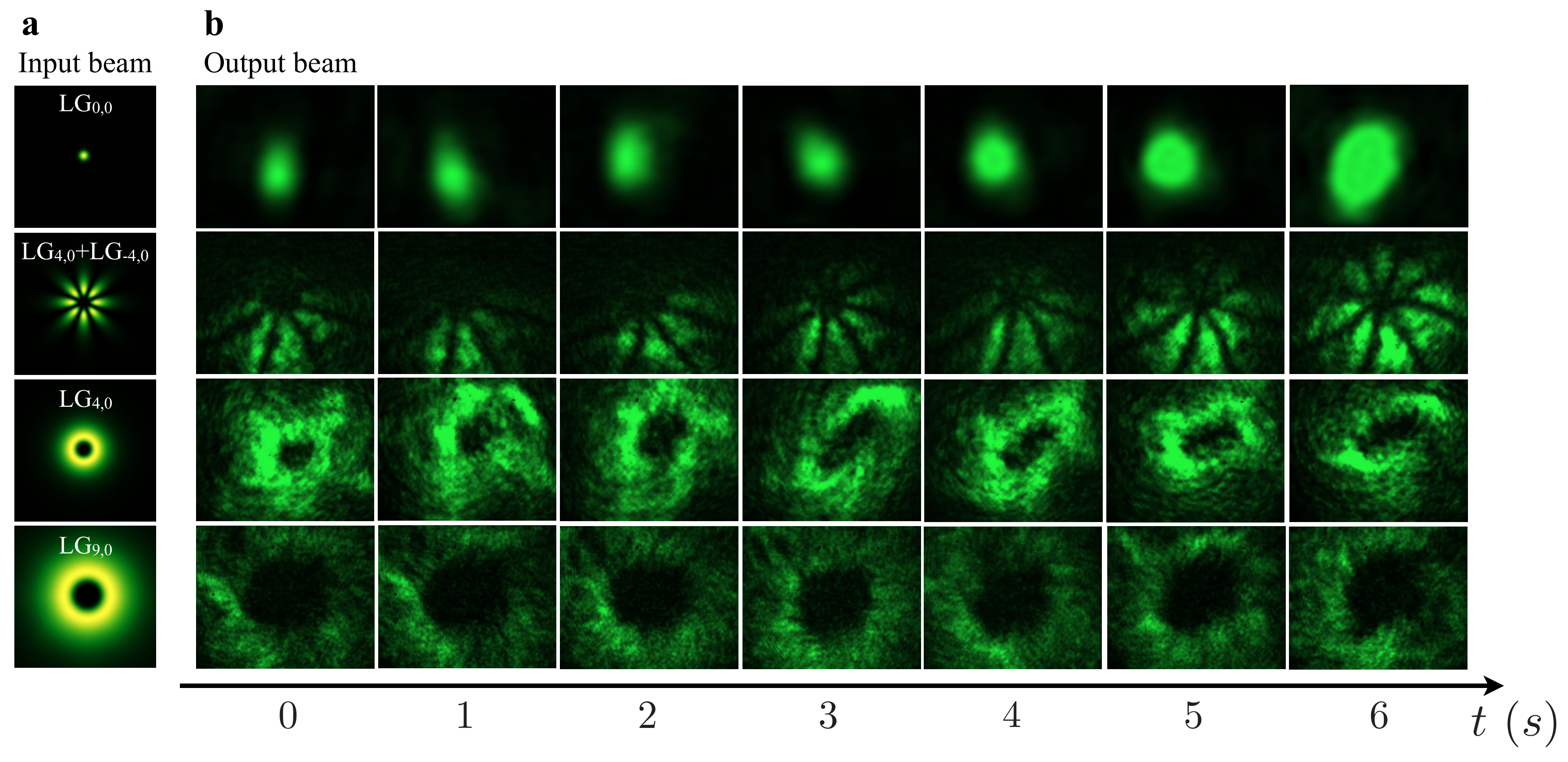}
		\caption[]{\textbf{Observed turbulence effects on spatial modes.} The images in the first, second, third and fourth rows correspond to $\text{LG}_{0,0}$, $(\text{LG}_{4,0}+\text{LG}_{-4,0})/\sqrt{2}$, $\text{LG}_{4,0}$, and $\text{LG}_{9,0}$, respectively. The original beam profile is shown in \textbf{a}, and the images after propagation through the underwater channel are shown in \textbf{b}. The images are taken over a 6 second interval. The exposure time for the last three rows are set to $108$~ms, while it is set to $20$~ms for the Gaussian beam. The inhomogeneity in the beams' intensity profiles are due primarily to Mie scattering from floating objects in the underwater channel. Lower and higher-order aberrations are manifested in the beam wandering (LG$_{0,0}$) and singularity splitting.}
		\label{fig:oam}
	\end{center}
\end{figure*}
In this experiment we prepared a Gaussian beam at the sender, and measure the wavefront at the receiver. The Gaussian beam should have a spherical phase due to divergence; thus, any variations from this can be attributed to turbulence introduced by the water. From the wavefront measurements, the turbulence can be expanded in terms of the Zernike coefficients, i.e. $\Phi (r,\phi) = \sum_j a_j Z_j(r,\phi)$. Here, $r$ and $\phi$ are the radial and azimuthal polar coordinates, respectively; $a_j$ are the Zernike expansion coefficients; $Z_j (r,\phi)= Z_n^m (r,\phi)$ are the Zernike polynomials depicted underneath the x-axis of figure 3a; $j=1+(n(n+2)+m)/2$ is the Noll index; and $n$ and $m$ are the radial and azimuthal indices, respectively. The values for the first 15 Zernike coefficients averaged from 30 wavefront measurements $\bar{a}_j$ are shown in Fig.~\ref{fig:wf}-\textbf{a} along with the reconstructed wavefront $\Phi$ from these values. In Fig.~\ref{fig:wf}-\textbf{b}, a sample of four of these individual wavefront measurements is shown. These wavefront measurements show that the beam experienced significant variation upon propagation through the turbulent channel. 

As stated before, the turbulence is also visible in the intensity profile of the beam at the receiver. It is easy to see tip-tilt aberrations from a Gaussian beam as it visibly drifts across the $x$ and $y$ axis of a camera. The higher order aberrations are often less visible. These aberrations do, however, show themselves very clearly in their effect on higher-order spatial modes. Specifically, the oblique and vertical astigmatism ($Z_2^{\pm2}(r,\phi)$) stretch OAM modes, giving them an elliptical shape, as well as splitting the singularity into lower topological charges. Intensity profiles of OAM and superposition modes are shown in Fig.~\ref{fig:oam} with consecutive images taken over a time of 6 seconds. The turbulence from the channel is very apparent in the wandering of the LG$_{0,0}$ mode, and the higher-order aberrations are shown most clearly in the stretching of the LG$_{0,4}$ mode. In addition to turbulence, all of the modes experience significant intensity fluctuations from changing levels of scattering as well as from objects floating into the beam's path. The latter is displayed clearly in the images of the petal beam, i.e. $(\text{LG}_{0,4}+\text{LG}_{0,-4})/\sqrt{2}$. As the correct measurement of spatial modes requires that the position and phase of the beam remain intact, it is clear that, even over a short propagation through water, active wavefront correction or the implementation of adaptive optics would be required to compensate for the aberrations.

\section{Conclusion}

We have shown that significant challenges present themselves in underwater communication. Despite low absorption from the water at blue-green wavelengths, the scattering from floating particles in the water can severely limit the achievable communication distance. Though scattering will impact the distance, we see that polarization states do maintain their integrity even after propagation through a very highly scattering channel. The second key challenge in an underwater optical channel is turbulence. This has the largest impact on communications using spatial modes. Through the $5.5$~m channel, the OAM modes experience aberrations which will result in errors in a communication protocol. The magnitude of these errors will need to be investigated in future work. Adaptive optics techniques will also need to be investigated to compensate for these errors and allow for communication using spatial modes.

\vspace{0.5 EM}
\noindent\textbf{Acknowledgments} This work was supported by Canada Research Chairs; Canada Foundation for Innovation (CFI); Canada First Research Excellence Fund (CFREF); Natural Sciences and Engineering Research Council of Canada (NSERC). We thank Denis Guay and the NRC for their help in designing and building the periscopes.


\end{document}